# Twist-tunable Polaritonic Nanoresonators in a van der Waals Crystal

O. G. Matveeva†, A. I. F. Tresguerres-Mata†, R. V. Kirtaev, K. V. Voronin, J. Taboada-Gutiérrez, C. Lanza-García, J. Duan, J. Martín-Sánchez, V. S. Volkov, P. Alonso-González*, A. Y. Nikitin*

† These authors contributed equally to this work.

Optical nanoresonators are fundamental building blocks in a number of nanotechnology applications (e.g. in spectroscopy) due to their ability to efficiently confine light at the nanoscale. Recently, nanoresonators based on the excitation of phonon polaritons (PhPs) – light coupled to lattice vibrations – in polar crystals (e.g. SiC, or h-BN) have attracted much attention due to their strong field confinement, high quality factors, and potential to enhance the photonic density of states at mid-infrared (IR) frequencies. Here, we go one step further by introducing PhPs nanoresonators that not only exhibit these extraordinary properties but also incorporate a new degree of freedom - twist tuning, i.e. the possibility to be spectrally controlled by a simple rotation. To that end, we both take advantage of the low-loss in-plane hyperbolic propagation of PhPs in the van der Waals crystal α-MoO$_3$, and realize a dielectric engineering of a pristine α-MoO$_3$ slab placed on top of metal ribbon grating, which preserves the high quality of the polaritonic resonances. By simple rotating the α-MoO$_3$ slab in the plane (from 0 to 45º), we demonstrate via far- and near-field measurements that the narrow polaritonic resonances (with quality factors Q up to 200) can be tuned in a broad range (up to 32 cm$^{-1}$, i.e up to ~ 6 times its full width at half maximum, FWHM ~ 5 cm$^{-1}$). Our results open the door to the development of tunable low-loss nanotechnologies at IR frequencies with application in sensing, emission or photodetection.

The excitation of PhPs in vdW crystals have recently emerged as an attractive strategy for manipulating IR light on deeply subwavelength scales[1-3], enabling fingerprint identification of organic molecules as well as strong coupling phenomena[4,5]. Some of the PhPs, such as those excited in α-MoO$_3$[6-11] or α-V$_2$O$_5$[12], present long lifetimes (up to a few ps) and in-plane hyperbolic propagation, being thus very appealing for the development of optical nanoresonators with ultra-low losses and potentially other unique properties. However, structuring of vdW crystals – needed for engineering nanoresonators – remains a challenging task, since the fabrication process dramatically degrades their optical properties, and thus the lifetime of PhPs[13]. Moreover, in contrast to plasmonic resonances in 2D materials, which are actively tunable via an external gate[14-16], tunability of PhPs nanoresonators has remained elusive. Here, we demonstrate PhPs nanoresonators fabricated in the van der Waals crystal α-MoO$_3$ that not only exhibit low losses, but also incorporate a unique twist tuning, i.e. the capability to be spectrally tuned by a simple rotation. The fabrication of these nanoresonators is based on placing a continuous (pristine) thin biaxial vdW crystal slab of α-MoO$_3$ on top of a grating formed by metal nanoribbons (see Methods). Remarkably, such design avoids any degradation of the optical properties of the crystal due to fabrication, maintaining the intrinsic low-loss and in-plane hyperbolic propagation of PhPs in α-MoO₃. The refractive indexes of the PhP modes supported by the α-MoO₃ slab above the air and metal regions are different, so that propagating PhPs modes can bounce back and forth between the boundaries of these regions (i.e., α-MoO₃/air and α-MoO₃/metal), eventually forming resonances. Most importantly, since the in-plane hyperbolicity of α-MoO₃, implies that different in-plane directions support PhPs with different momenta, the mutual orientation of the crystal axes and the metal grating can serve as a tuning knob, in analogy to recent demonstrations of twistoptics in α-MoO₃ bilayers[17-20]. Thus, by twisting the α-MoO₃ slab with an in-plane angle with respect to the metal ribbons, we demonstrate a broad tunability of the PhPs nanoresonators, as revealed by both far- and near-field techniques: Fourier-transform IR spectroscopy (FTIR) and scattering-type scanning near-field optical microscopy (s-SNOM), respectively. Furthermore, with the help of theoretical analysis, we are able to interpret and disentangle the different resonant PhPs modes observed in the experimental data.

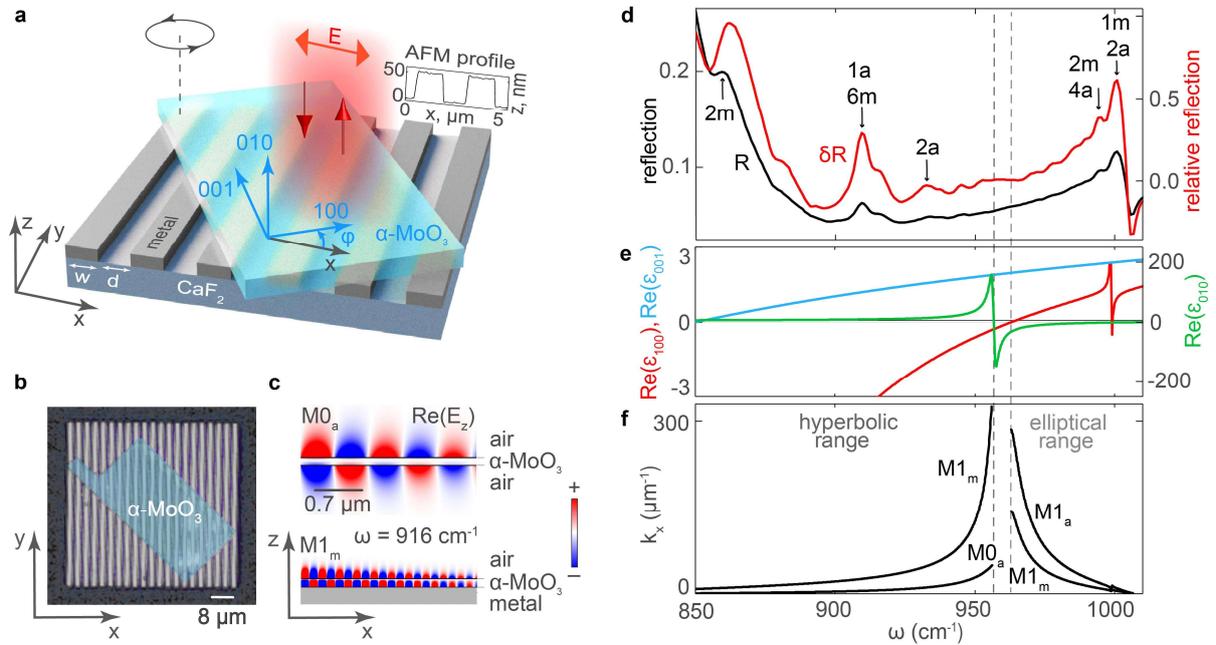

**Figure 1.** PhPs nanoresonators in α-MoO$_3$ defined by placing a pristine α-MoO$_3$ slab on top of metal ribbons. a) Schematics of the studied structure that allows defining the nanoresonators by "dielectric engineering" and controlling them by a twist angle, $\varphi$. The inset shows the AFM profile of the metal grating. b) False-colors optical image (top view) of the sample consisting of a 110 nm-thick α-MoO$_3$ slab placed on a 50 nm-thick metal grating. c) Simulated field distributions of the M0 and M1 PhPs modes in the α-MoO$_3$/air (top) and α-MoO$_3$/metal (bottom) regions. d) Far-field spectra of the PhPs nanoresonators (raw and relative reflection spectra are displayed in black and red curves, respectively) for $\varphi = 0°$. The subscripts "a" and "m" indicate the PhPs resonances originated in the α-MoO$_3$/air and α-MoO$_3$/metal regions, respectively. e) Real part of the dielectric permittivity tensor components as a function of $\omega$. f) Calculated dispersion of the M1 and M0 modes shown in (c).

Figure 1a illustrates schematically the heterostructure designed to define "dielectric-engineered" PhPs nanoresonators in our work. It consists of an α-MoO$_3$ slab placed on top of a periodic array of metal ribbons fabricated on a IR-transparent CaF$_2$ substrate (see the AFM profile in the inset to Figure 1a and the optical image in Figure 1b), similarly to the concept of polaritonic nanoresonators based on in-plane isotropic slabs of h-BN and graphene layers[21-23]. The α-MoO$_3$ slab is twisted with respect to the direction across the ribbons (x-direction) by an arbitrary angle $\varphi$. The metal arrays (with ribbon widths $w = 1480$ nm and a separation distance $d = 1230$ nm) have been designed so that the heterostructure exhibits PhPs resonances at mid-IR frequencies (in the range $\omega = 850 - 1010$ cm$^{-1}$). The reflection spectra taken by FTIR upon incident illumination polarized across the ribbons and along the α-MoO$_3$ [100] crystal direction ($\varphi = 0°$) is illustrated in Figure 1d (black curve). To better recognize the resonance spectral features, we normalize the reflection coefficient, $R$, to its moving average, $\bar{R}$, obtaining the relative reflection $\delta R = (R - \bar{R})/\bar{R}$ (Figure 1d, red curve), which clearly demonstrates a set of sharp resonance peaks. Particularly, these peaks appear within the RBs of the α-MoO$_3$ crystal (Figure 1e), defined in the spectral ranges $\omega = 850 - 956.8$ cm$^{-1}$, where Re($\varepsilon_{100}$) < 0, and $\omega = 963 - 1006$ cm$^{-1}$, where Re($\varepsilon_{010}$) < 0, exhibiting Q as large as 180 at 916 cm$^{-1}$ and 200 at 993 cm$^{-1}$ (see Supporting Note V). In the lower (upper) RB PhPs exhibit hyperbolic (elliptic) dispersion[24], composing a set of in-plane anisotropic electromagnetic modes propagating along the crystal slab. Importantly, the PhPs modes in the region α-MoO$_3$/air (M$l$$_a$) are expected to have a different refractive index (momentum) to those in the region α-MoO$_3$/metal (M$l$$_m$), even for the same mode number, $l$, the latter representing the quantization of the PhP's electric field in the z-direction[9,25]. This is corroborated by plotting the dispersions of the fundamental modes in both regions within the hyperbolic and elliptic RBs (M0$_a$, M1$_a$ and M1$_m$ in Figure 1f, with the hyperbolic modes corresponding to the α-MoO$_3$ [100] crystal direction). Clearly, the M1$_m$ mode exhibits a much shorter wavelength (larger momentum), together with a stronger field confinement to the faces of the α-MoO$_3$ slab (Figure 1c). We thus assume that the resonant features seen in the spectra of Figure 1d correspond to Fabry-Pérot resonances (FPR) arising from multiple reflections of the PhP waveguiding modes due to the mode's refractive index steps defined along the x-axis.

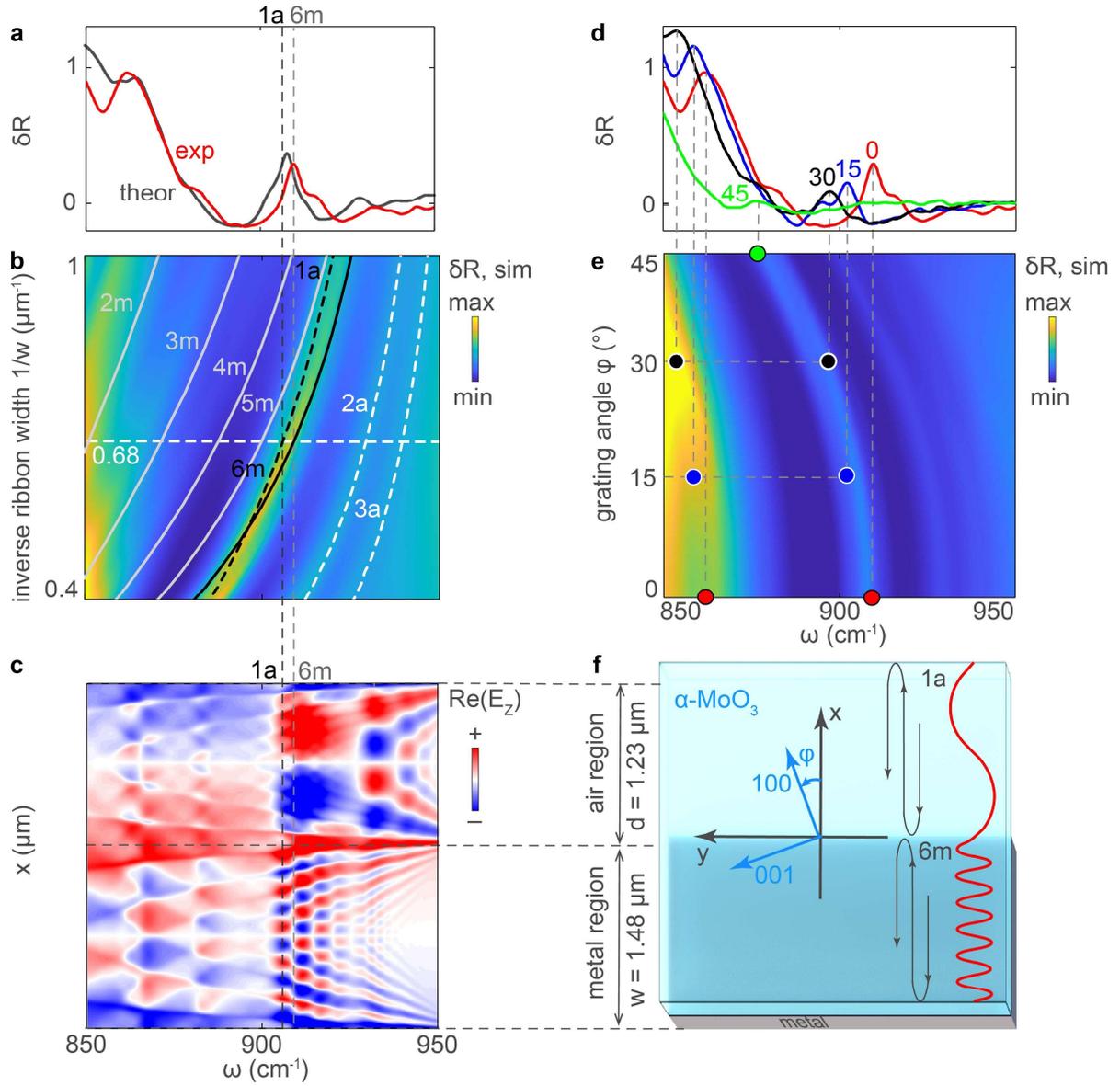

**Figure 2.** Analysis of the PhPs resonances and their twist tuning. a) Measured and simulated relative reflection spectra, $\delta R$ (solid red and dash grey curves, respectively), for a ribbon width $w = 1480$ nm, a separation distance $d = 1230$ nm and a twist angle $\varphi = 0°$. b) Simulated $\delta R$ as a function of frequency, $\omega$, and inverse ribbon width, $1/w$, for a fixed ratio $w/d$. The FBR conditions for PhP in α-MoO$_3$ above the air and metal regions are traced by the white solid and dashed lines, respectively. c) Simulated field distributions of the M0 and M1 PhP modes in the

α-MoO3/air and α-MoO3/meal regions, respectively, as a function of $\omega$ and the $x$ coordinate (short axis of the ribbons). d) Measured $\delta R$ spectra for twist angles $\varphi = 0, 15, 30$ and $45°$ (red, blue, black and green, respectively). e) Simulated relative reflection as a function of $\omega$ and $\varphi$. f) Top view of one lattice unit cell.

To corroborate our assumption on the physical origin of the PhPs resonances and explain the reflection spectra obtained experimentally, we perform a theoretical analysis based on full-wave simulations of a structure mimicking the fabricated nanoresonators. The simulated relative reflection spectrum (grey curve in Figure 2a) shows a good agreement with the experiment (red curve in Figure 2a) for the case of $w = 1480$ nm and $d = 1230$ nm. We, therefore, extend the theoretical study for a broad range of $w$ and $d$ in Figure 2b, where $\delta R$, represented by a colorplot, shows a set of maxima that clearly depend on the ribbon width or the distance between the ribbons. These resonances presumably coincide with an anisotropic FPRs condition, which can be written in a simple form according to the PhP mode phase matching as $a \cdot k_M + \phi_M = \pi n$, analogously to that for in-plane isotropic FPRs in graphene ribbons[23,26]. In our case, $a$ coincides with either $w$ or $d$, depending on the region in the α-MoO3 slab where the nanoresonators are defined (above the metal or air, respectively), $k_M$ and $\phi_M$ are the momentum and reflection phase of the $M$th mode, respectively, and $n$ is the integer numerating the FPRs (quantization of the $M$th PhP mode in the x-direction), indicating how many wavelengths of the mode fit across either the distance $w$ or $d$. This numeration includes only the "bright" FPRs, corresponding to the non-zero overlap between the illuminating wave and the PhPs in a FPR cavity[26]. Plugging the $k_M$ from the dispersion relation of PhP modes in a biaxial crystal slab[9,25] into the FPR condition, we obtain a set of curves for each of the modes (rendered on top of the colorplot in Figure 1b). These curves match the maxima of $\delta R$, thus confirming the Fabry-Pérot origin of the resonances in our polaritonic structures. The FPR can also be recognized in the calculated spatial distribution of the vertical electric field above the unit cell of the periodic structure (see schematics in Figure 2f), represented as a function of frequency $\omega$ and coordinate $x$ in Figure 2c. Indeed, at the peak frequencies the number of field oscillations across the ribbon matches with the quantization number $n$ in the FPR condition, both for the PhP modes in the α-MoO$_3$/air regions (marked with "$n$a") and in the α-MoO$_3$/metal regions (marked with "$n$m").

Importantly, the spectral position of the PhP resonances strongly depends on the twist angle $\varphi$ between the α-MoO$_3$ slab and the metallic ribbons, as clearly observed in the reflection spectra of Figure 2d. In particular, the main resonant peak (1a) shifts to lower frequencies (from ~ 909.5 to ~ 877 cm$^{-1}$) with increasing $\varphi$ (the increasing signal-to-noise ratio enables determining the resonant features for $\varphi$ up to 45º), which can also be seen in our theoretical calculations (Figure 2e). Note the much higher Q (see Supporting Note V) of our twist-tunable PhP resonators (by more than an order of magnitude, due to both the absence of the electron scattering and avoiding the direct crystal structuring) compared to their plasmonic counterpart reported at THz frequencies in in-plane anisotropic WTe$_2$ crystals[27].

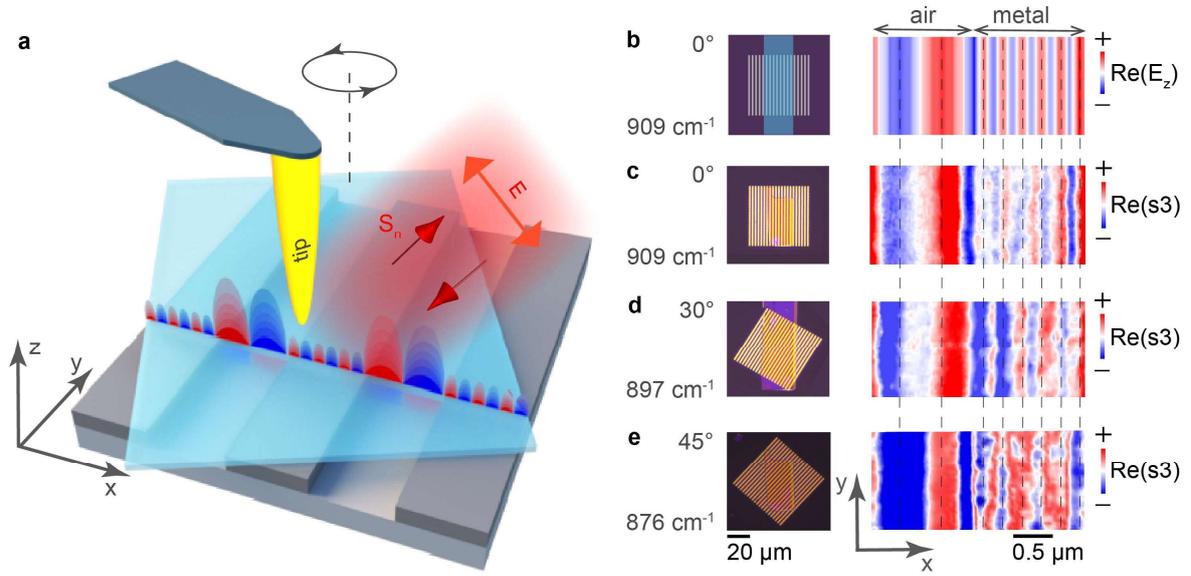

**Figure 3.** Near-field measurements of twist-tunable PhPs nanoresonators. a) Schematic of the near-field measurements by s-SNOM, in which a metal tip is illuminated by p-polarized mid-IR light and both the amplitude and phase of the tip-scattered field are recorded as a function of the tip position. b) Simulated field distribution, Re($E_z$), in the α-MoO$_3$/air and α-MoO$_3$/metal regions. c-e) Experimental near-field images, Re($s3$), of one of the unit cells of the heterostructure for the twist angles $\varphi = 0, 30,$ and 45º, respectively, taken at the resonant frequencies.

To further corroborate our results and the structure of the electric field in the anisotropic PhP nanoresonators, we perform near-field measurements by s-SNOM (see Methods). By scanning the α-MoO$_3$ slab with the s-SNOM tip (Figure 3a), we record the electromagnetic signal as a function of the tip position, composing the near-field images (Figures 3c-e). At a frequency $\omega = 909$ cm$^{-1}$ (Figure 3c), corresponding to the 1a and 6m resonances in the far-field spectra (Figure 2a), we observe signal fringes across the ribbons, indicating the excitation of PhPs. The much larger distance between the red and blue fringes in the region α-MoO$_3$/air (1 oscillation) compared to the region α-MoO₃/metal (6 oscillations) is consistent with the much longer PhP wavelength (see the simulated PhP field distribution along the black and grey vertical dashed lines, "1a, 6m", in Figure 2c), as expected from their dispersion relation (Figure 1f). For a better visualization, we represent the simulated $\text{Re}(E_z)$ as a function of the in-plane coordinates $x, y$ in Figure 3b, which perfectly matches with the near-field image in Figure 3c. From these observations, we can thus conclude that: (i) the s-SNOM tip-scattered signal represents a measure of $\text{Re}(E_z)$ above the slab[13,28] and (ii) the visualized near-fields at the resonant peaks of the far-field spectra are clearly consistent with the FPR model. Remarkably, changing both the measuring frequency and the twist angle, following the experimental data shown in Figure 2d, the number of oscillations in the near-field images keeps constant (the agreement between the periods of the near-field signal oscillations across the ribbons is shown in Figures 3c-e by the vertical dashed eye-guides). Namely, the calculated PhPs wavelength in the area above the air gap (1a resonance) is equal to $\lambda_p = 1.29$ μm, approximately fitting the air gap width (1.23 μm), while in the area above the metal (6m resonance) $\lambda_p = 249$ nm, thus being consistent with the ribbon width divided by 6 (247 nm). Therefore, our near-field measurements reconfirm the emergence of FPR of the same order for the chosen parameters. On the other hand, although similar oscillations are observed in the near-field images, the FPR originates from PhPs at different points in the dispersion surface (oscillating at different frequencies and propagating along different directions).

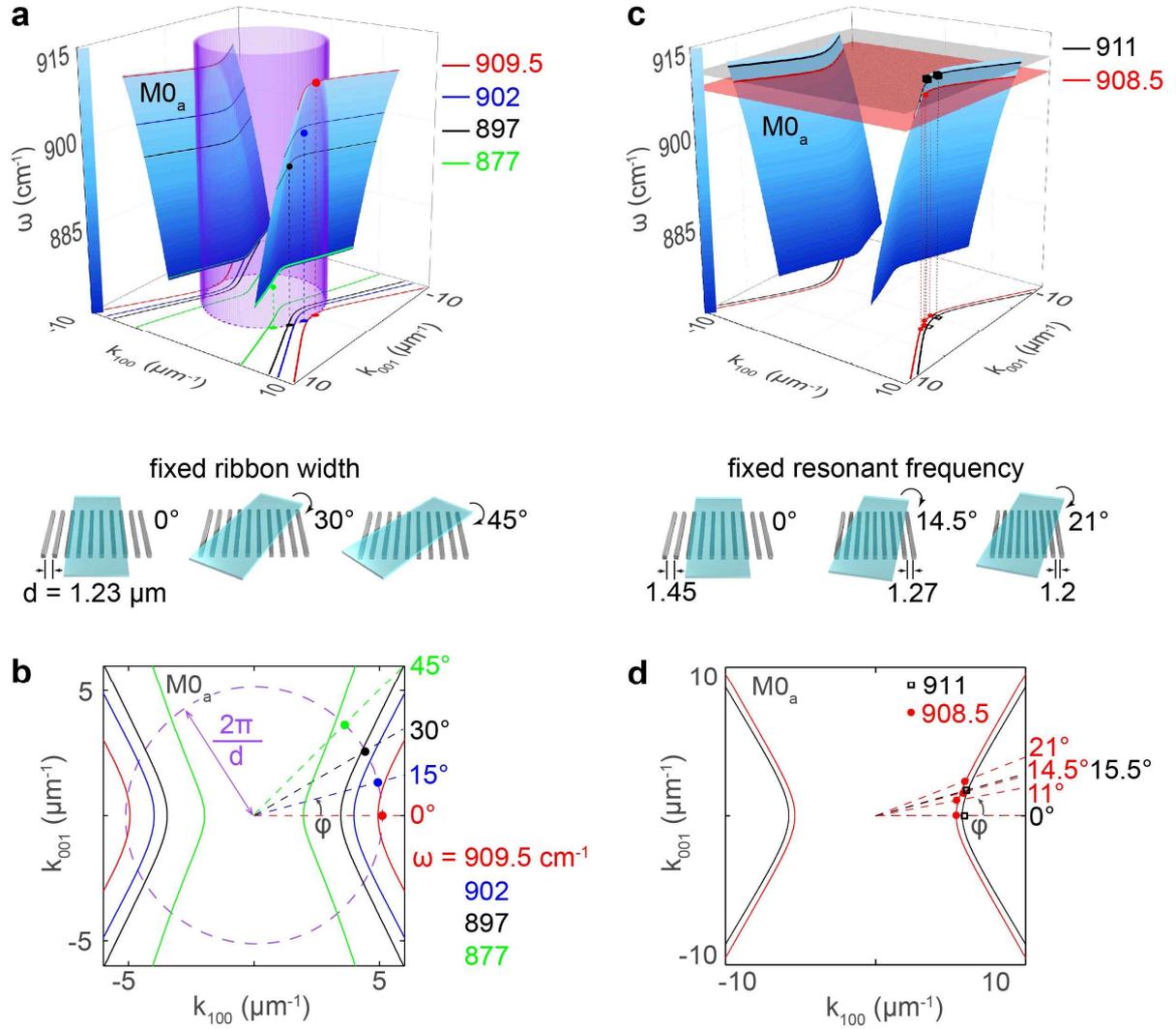

**Figure 4.** Probing the dispersion surface of the M0$_a$ PhPs mode. a) Analytical dispersion surface of the M0$_a$ PhPs mode (blue hyperboloid) when crossed by a surface (purple cylinder) representing a constant momentum. The color dots represent the positions of the measured resonant peaks (from Figure 2d) for the same metal grating twisted at different angles, $\varphi$. b,d) Isofrequency curves for the M0$_a$ mode at different frequencies $\omega$. c) Analogous dispersion surface as in (a), but crossed by planes of fixed frequencies ($\omega$ = 908.5, and 911 cm$^{-1}$). The points mark the peak positions for metal lattices with different widths of the air gaps $d$ and twist angles $\varphi$. The thicknesses of the flakes in (a,b) and (c,d) are 110 and 127.5 nm, respectively.

In order to illustrate "probing" of the PhP modes in our nanoresonators at different points of the dispersion surface, we mark the positions of the main peaks in the normalized reflection spectra (Figure 2d) directly on the hyperboloids representing the dispersion of the M0a PhP mode (Figure 4a). The experimentally-measured positions of the resonances (green, black, blue and red points in Figure 4a) can be represented in polar coordinates by the frequency, $\omega$, of the resonant peak (z coordinate), the inverse air gap width, $2\pi/d$ (radial coordinate), and the twist angle, $\varphi$. The selection of the PhPs momenta for the first-order FPR at any $\varphi$ can be visualized by a cylinder with radius, $2\pi/d$, passing through the dispersion surface (Figure 4a). Crossings between the latter and the cylinder give the approximate positions of the FPR, which can be more clearly seen in the $k_x, k_y$ plane (Figure 4b), where the isofrequency curves (IFCs) for each resonant frequency are shown. The experimental points fall very close to the crossings between the circle and the IFCs, thus reconfirming our initial assumption on the Fabry-Pérot origin of the resonances.

Alternatively, the PhP dispersion surface can also be probed by cutting it with planes of constant frequencies, $\omega$. To that end, we performed an additional set of far-field measurements for metal gratings with different air gaps, $d$, and twist angles, $\varphi$, all of them matching the same FPR (see schematics in Figure 4). Namely, we designed our structures to exhibit the "1a" FPR at two fixed frequencies: 908.5 cm$^{-1}$ and 911 cm$^{-1}$ (see measurements in the Supporting Note I). The positions of the measured resonant peaks perfectly match with the crossing points between the calculated IFCs and the radial lines from the origin at an angle $\varphi$ (dashed lines in Figure 4d), as seen in Figures 4c,d for both frequencies. Note that a similar analysis can be also easily performed for the FPRs of other orders, and particularly for the FPRs composed by the PhP modes above the metallic region, M1$_m$ (see Supporting Note III). This analysis is equally suitable for the upper RB, i.e. for the elliptic regime (see Supporting Note IV).

To summarize, we introduce Fabry-Pérot nanoresonators that exhibit a unique tuning by a simple rotation of the host crystal. This is achieved by fabricating heterostructures composed of metal gratings and twisted vdW crystal slabs supporting in-plane anisotropic PhPs ($\alpha$-MoO$_3$). In contrast to conventional Fabry-Pérot polaritonic resonators, in which the reflecting boundaries are typically fabricated by etching the polaritonic material, the design of our tunable Fabry-Pérot nanoresonators allows preserving the crystalline properties of the slabs and thus obtaining high Q. Interestingly, the resulting FPRs, visualized in real space by s-SNOM, allow reconstructing the three-dimensional anisotropic dispersion surfaces of the PhPs from data collected at either fixed momentum or fixed frequency. Such reconstruction of the polaritons dispersion opens the door to an alternative characterization of novel vdW materials, particularly for a more reliable determination of their optical properties. From an application point of view, our nanoresonators can be very appealing for tunable mid-IR narrow-resonance sensors, and in particular for molecular bar-coding[29].

METHODS

**Fabrication of metal ribbons.** Arrays of ribbons of both Au and Al were fabricated by electron beam lithography. In the case of Al ribbons, first, a 70 nm-thick Al layer was deposited by electron beam evaporation on a CaF$_2$ substrate. Subsequently, a negative photoresist layer (MA-N2401) was spin coated (3000rpm -> 90 nm) on top of the Al layer for the definition of the ribbon pattern by electron beam lithography (50keV, 200pA, dose 280 uC/cm$^2$). A high-resolution developer AZ726 was used for the lift-off. By reactive-ion etching of Al in BCL3/Cl2 plasma (pressure 40mT, RIE power 100W) and removing the photoresist in oxygen plasma, we got the set of 50 μm (length) × 60 μm (width) × 70 nm (height) gratings with different ribbon widths. In the case of Au ribbons, high-resolution electron beam lithography was employed under 100kV and 100pA. First, the samples were coated with a positive resist layer (PMMA). To dissolve the exposed areas, a conventional high-resolution developer (1:3 MIBK: IPA) was used. Afterwards, 5 nm of Cr and 50 nm of Au were evaporated. Finally, the lift-off was performed to define the 50 μm (length) x 50 μm (width) x 50 nm (height) Au grating with ribbons of 1.48 μm width.

We used the set of Al gratings exclusively for measurement presented in the Figures 4c,d. For all other measurements we used the Au grating.

**Fabrication and In-plane Twist of α-MoO$_3$ Flakes.** α-MoO$_3$ flakes were mechanically exfoliated using a Nitto tape (Nitto Denko Co., SPV 224P) from commercial α-MoO$_3$ bulk crystals (Alfa Aesar). A second exfoliation was performed from the tape to transparent polydimethylsiloxane (PDMS) in order to thin them down. The flakes were examined with an optical microscope in order to select homogeneous pieces with the desired thicknesses (110 nm and 127.5 nm) and large surface areas, which were picked up with the help of polycarbonate (PC). The dry transfer technique was used, with the help of a micromanipulator, the flakes were precisely aligned and twisted on top of the metal grating. Transferring was carried out by heating up to 250°C in order to liquate the PC. The PC was removed with chloroform at 100°C releasing the flake.

**Fourier-Transform Infrared Spectroscopy.** The far-field optical response of our α-MoO$_3$ nanoresonators was characterized by Fourier-Transform Infrared Spectroscopy (FTIR) using a Varian 620-IR microscope coupled to a Varian 670-IR spectrometer, supplied with a broadband mercury cadmium telluride (MCT) detector (400-6000 cm$^{-1}$). The spectra were collected with a 2 cm$^{-1}$ spectral resolution and the spatial resolution was adjusted to the size of the metal arrays. The infrared radiation from the thermal source (normal incidence) was linearly polarized employing a wire grid polarizer. CaF$_2$ substrates were used due to their transparency in the spectral range under study.

**Scattering-Scanning Near Field Optical Microscopy.** Near-field imaging measurements were performed by employing a commercial scattering-type Scanning Near Field Optical Microscope (s-SNOM) from Neaspec GmbH, equipped with a quantum cascade laser from Daylight Solutions (890-1140 cm$^{-1}$). Metal-coated (Pt/Ir) atomic force microscopy (AFM) tips (ARROW-NCPt-50, Nanoworld) at a tapping frequency $\Omega \sim 280$ kHz and an oscillation amplitude $\sim 100$ nm were used as source and probe of polaritonic excitations. The light scattered by the tip was focused by a parabolic mirror into an infrared detector (Kolmar Technologies). Demodulation of the detected signals to the 3rd harmonic of the tip frequency ($s_3$) was carried out for background suppression. A pseudo-heterodyne interferometric method was employed to independently extract both amplitude and phase signals.

**Full-wave Numerical Simulations.** Two types of full-wave numerical simulations were performed using COMSOL software, based on the finite-element method in the frequency domain. In both cases, the structure was composed of the α-MoO$_3$ flake on top of the array of metal ribbons placed on the semi-infinite CaF$_2$ substrate. The first type of simulation was based on the far-field illumination of the structure by a normally-incident plane wave polarized across the ribbons. The ratio of the ribbon to the air gap widths was fixed constant $w/d = 1480/1230$ (the experimental measurements for the array of ribbons with such parameters were performed in Figures 1d, 2a and 3b-f). The reflection coefficient and the spatial distribution of the vertical electric field, $E_z$, above the α-MoO$_3$ slab have been extracted from the simulations (see Fig 2a-e, Fig 3b-e). In the second type of simulation, we have used the quasi-normal eigenmode analysis in order to find the electric field distribution of the modes M0 and M1 (see Fig 1c).

In the analytical analysis of the FPR (Figure 2b and Figure 4), the reflection phase of the PhPs modes, $\phi_M$, in the α-MoO$_3$/metal and α-MoO$_3$/air regions is taken 0 and $\pi$, respectively.

ASSOCIATED CONTENT

**Supporting Information.**

The Supporting Information is available free of charge in Supporting_information.pdf .

AUTHOR INFORMATION

**Corresponding Authors**

**A. Y. Nikitin** − *Donostia International Physics Center (DIPC), 20018 Donostia-San Sebastián, Spain; IKERBASQUE, Basque Foundation for Science, 48013 Bilbao, Spain; orcid.org/0000-0002-2327-0164; Email: alexey@dipc.org.*

**P. Alonso-González** − *Department of Physics, University of Oviedo, 33006 Oviedo, Spain; Center of Research on Nanomaterials and Nanotechnology CINN (CSIC-Universidad de Oviedo), 33940 El Entrego, Spain; E-mail: pabloalonso@uniovi.es*

**Authors**

**O. G. Matveeva** − *Center for Photonics and 2D Materials, Moscow Institute of Physics and Technology, 141700 Dolgoprudny, Russia*

**A. I. F. Tresguerres-Mata** − *Department of Physics, University of Oviedo, 33006 Oviedo, Spain*

**R. V. Kirtaev** − *Center for Photonics and 2D Materials, Moscow Institute of Physics and Technology, 141700 Dolgoprudny, Russia*

**K. V. Voronin** − *Center for Photonics and 2D Materials, Moscow Institute of Physics and Technology, 141700 Dolgoprudny, Russia; Donostia International Physics Center (DIPC), 20018 Donostia/San Sebastián, Spain*

**J. Taboada-Gutiérrez** − *Department of Physics, University of Oviedo, 33006 Oviedo, Spain; Center of Research on Nanomaterials and Nanotechnology CINN (CSIC-Universidad de Oviedo), 33940 El Entrego, Spain*

**C. Lanza-García** − *Department of Physics, University of Oviedo, 33006 Oviedo, Spain*

**J. Duan** − *Department of Physics, University of Oviedo, 33006 Oviedo, Spain; Center of Research on Nanomaterials and Nanotechnology CINN (CSIC-Universidad de Oviedo), 33940 El Entrego, Spain*

**J. Martín-Sánchez** − *Department of Physics, University of Oviedo, 33006 Oviedo, Spain; Center of Research on Nanomaterials and Nanotechnology CINN (CSIC-Universidad de Oviedo), 33940 El Entrego, Spain*

**V. S. Volkov** – *Center for Photonics and 2D Materials, Moscow Institute of Physics and Technology, 141700 Dolgoprudny, Russia; GrapheneTek, Skolkovo Innovation Center, 143026 Moscow, Russia*

**Author Contributions**

A.Y.N. and P.A.-G. conceived the study and supervised the work. R.V.K. performed the fabrication of the Al ribbons under the supervision of V.S.V. A.I.F.T.-M. performed the far- and near-field experiments with the help of J.D. A.I.F.T.-M. carried out the sample preparation with the help of J.T.-G., C.L.-G. and J.M.-S. O.G.M. performed data analysis and simulations with the input from K.V.V. A.Y.N. and O.G.M. wrote the manuscript with input from A.I.F.T.-M, J. D, and P.A.-G. All authors contributed to the scientific discussion and manuscript revisions.

ACKNOWLEDGMENT

The authors acknowledge Irene Dolado López for her advice and assistance in the Au gratings fabrication. O.G.M., R.V.K., K.V.V. and V.S.V. acknowledge financial support from the Ministry of Science and Higher Education of the Russian Federation (Agreement No. 075-15-2021-606). A.I.F.T.-M. and J.T.-G. acknowledge support through the Severo Ochoa program from the Government of the Principality of Asturias (nos. PA-21-PF-BP20-117 and PA-18-PF-BP17-126, respectively). J.M.-S. acknowledges financial support through the Ramón y Cajal program from the Government of Spain (RYC2018-026196-I). P.A.-G. acknowledges support from the European Research Council under starting grant no. 715496, 2DNANOPTICA and the Spanish Ministry of Science and Innovation (State Plan for Scientific and Technical Research and Innovation grant number PID2019-111156GB-I00). A.Y.N. acknowledges the Spanish Ministry of Science and Innovation (grants MAT201788358-C3-3-R and PID2020-115221GB-C42) and the Basque Department of Education (grant PIBA-2020-1-0014).

Supporting Information for "Twist-tunable Polaritonic Nanoresonators in a van der Waals Crystal"

**Table of Contents**

In this Supporting Information we will consider the Fabry- Pérot resonances (FBRs) with their origin in: i) the M1 PhP mode propagating in the $\alpha$-MoO$_3$ flake above the metal region in the hyperbolic frequency range, and ii) the M1 PhP mode propagating in the flake above both the metal ribbon and the air gap regions in the elliptic frequency range.

## I.   Far-field Characterization

The far-field characterization of our samples was carried out by performing FTIR measurements on an $\alpha$-MoO$_3$ flake with thickness $t = 110$ nm placed on top of Au ribbons with a width $w = 1.48$ µm and a separation $d = 1.23$ µm (FigureS1a). Within the hyperbolic frequency range we have identified the FPRs 1a and 6m, matching the same resonance frequency ($\omega = 909.5$ cm$^{-1}$, see Figure1d of the main text). In contrast, in the elliptic range we have identified the resonances 2m and 4a close to $\omega = 995$ cm$^{-1}$ (Figure1d in the main text). The analysis of the resonances 2m and 4a for different twisting angles $\varphi$ is presented in FigureS3d of Section II.

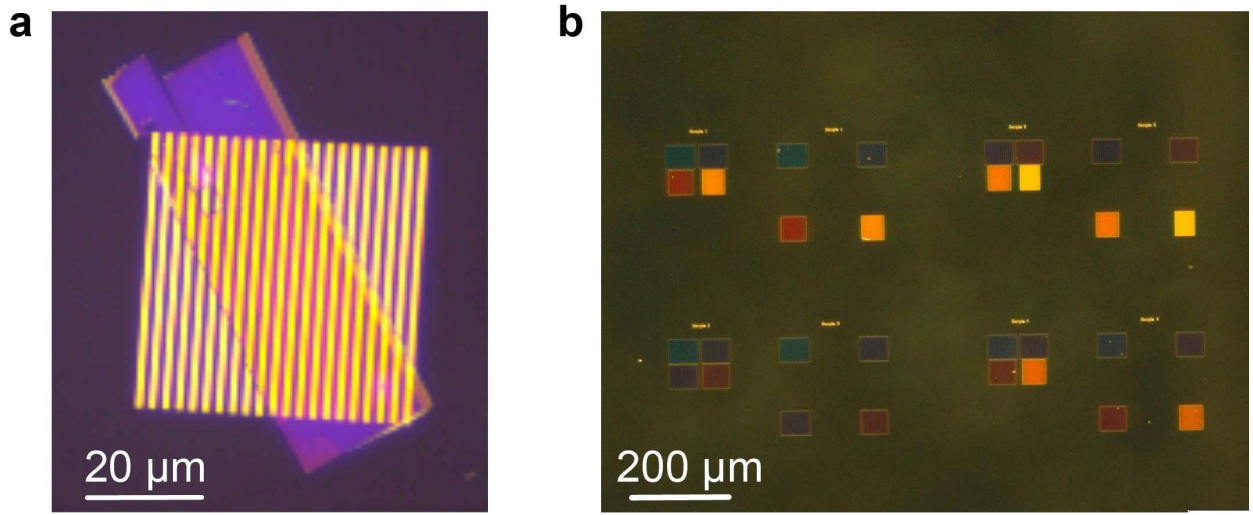

**Figure S1.** Optical images of the metal gratings. a) Au grating with a ribbon width $w = 1.48$ µm and a separation (air gaps) $d = 1.23$ µm. The $\alpha$–MoO$_3$ flake placed on top of them has a thickness $t = 110$ nm. b) Sets of Al gratings with different ribbon widths and air gaps. The substrate is in all cases CaF$_2$.

Additionally, we fabricated Al gratings (as the type of metal does not affect our results in the mid-IR, in some cases we use Al instead of Au for convenience) with different periods (FigureS1b). In this case, the $\alpha$–MoO$_3$ flake placed on top had a thickness $t = 127.5$ nm. By varying the twisting angle of the flake with respect to the ribbons axes, we observe the same resonant frequencies as when using Au in both hyperbolic and elliptic frequency ranges (FigureS2a,b, and FigureS2c, respectively). The FPRs modes 1a and 6m in the hyperbolic range appear at $\omega = 908.5$ cm$^{-1}$ (Figure S2a) and $\omega = 911$ cm$^{-1}$ (Figure S2b), while the FBRs 2m and 4a in the elliptic range appear at $\omega = 991$ cm$^{-1}$ (FigureS2c). The ribbons and air gaps widths were measured by AFM.

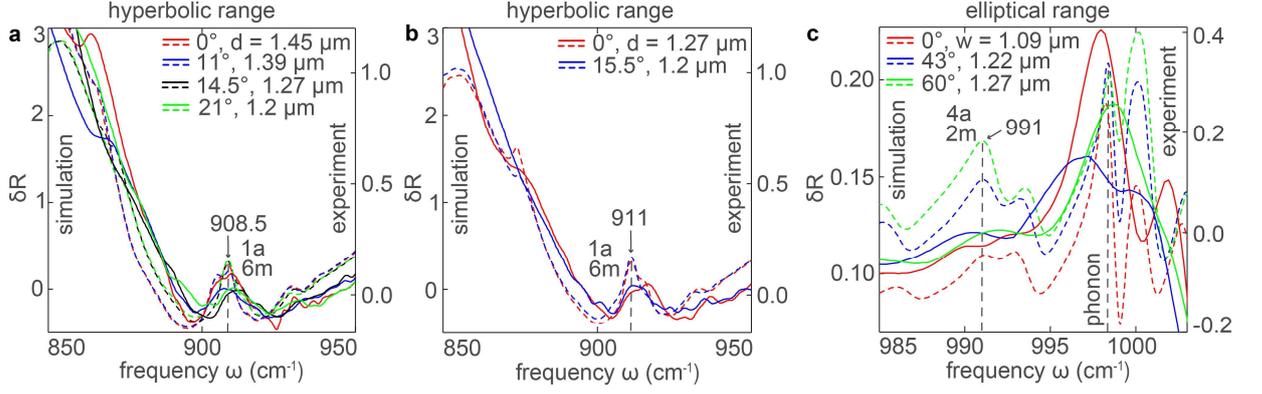

**Figure S2.** Far-field spectra of PhP nanoresonators in the case of using Al gratings. a) Experimental and simulated relative reflection spectra in the hyperbolic range for different air gap widths, $d$, and flake rotation angles, $\varphi$, showing resonances (FPRs 1a and 6m for the M0 and M1$_m$ PhP modes, respectively) at the same frequency, $\omega = 908.5$ cm$^{-1}$. b) Analogous spectra as in (a) for gratings with a resonant frequency $\omega = 911$ cm$^{-1}$. c) Experimental and simulated relative reflection spectra in the elliptic range for nanoresonators defined on gratings with different ribbon widths $w$ and twisting angles $\varphi$. The nanoresonators show maximum of their relative reflection at the same frequency $\omega = 991$ cm$^{-1}$ (resonances 2m and 4a for the M1$_m$ and M1$_a$ PhP modes, respectively). Solid and dash lines in (a-c) represent experimental and simulated data, respectively. The thickness of the α−MoO$_3$ flake in was 127.5 nm.

Figs.S2a-c show the relative reflection spectra for an α-MoO$_3$ flake placed on top of Al gratings with different air gaps (Figs.S2a,b) and ribbon widths $w$ (Figs.S2c). For all the gratings the flake was rotated by a given angle $\varphi$ such that the resonances 1a and 6m (Figs.S2a,b) as well as the resonances 2m and 4a (Figs.S2c) take place at the same frequency.

## II. Analysis of the Anisotropic PhP Resonances in the Elliptic α-MoO$_3$ Frequency Range

The FTIR relative reflection spectra (red dots in FigureS3a), hereinafter referred to as $\delta R$, of nanoresonators in the elliptic α-MoO$_3$ frequency range was measured for a 110nm-thick α-MoO$_3$ flake placed on top of an array of Au-ribbons (with a period of 2.71 µm and a ribbon width of 1.48 µm, corresponding to $d = 1.23$ µm) at $\varphi = 0°$. For comparison, full-wave numerical simulations of $\delta R$ (black curve in FigureS3a) were carried out mimicking the experiment (in both cases using plane wave illumination polarized across the ribbons). Although the agreement between experiment and theory is reasonably good, the spectral resolution in our FTIR measurements (2 cm$^{-1}$) is too low to capture the TO-phonon along the [100] crystal direction at $\omega = 998.7$ cm$^{-1}$. The colorplot in FigureS3b shows $\delta R$ as a function of frequency and inversed ribbon width (while the ratio between the ribbon width and the width of the air gap is fixed to 1.48/1.23).

In our system, we consider the FPRs in both regions of the flake (above the metal and above air) as independent (i.e. completely neglecting any coupling between them). In this approximation, we can write the phase matching conditions[1]:

$$k_a g + \Phi_a = \pi n_a \tag{1}$$

$$k_\mathrm{m} w + \Phi_\mathrm{m} = \pi n\mathrm{m}, \qquad (2)$$

where $k_a$ and $k_\mathrm{m}$ are the wavevectors in the regions above the air gaps and the metal ribbons, $\Phi_\mathrm{a}$ and $\Phi_\mathrm{m}$ are the reflection phases from the air/metal and metal/air boundaries, and $n$a and $n$m represent the number of polariton wavelengths fitting in the areas above the air gaps and metal ribbons, respectively. We take the wavevectors in Eqs. (1,2) from the analytical equation for the dispersion of anisotropic modes M$l$ in a thin biaxial slab of thickness $d$, propagating in the plane at an angle $\varphi$ with respect to the [100] crystallographic direction[2]:

$$k(\omega) = \frac{\rho}{k_0 d}\left[\arctan\left(\frac{\varepsilon_1 \rho}{\varepsilon_z}\right) + \arctan\left(\frac{\varepsilon_1 \rho}{\varepsilon_z}\right) + \pi l\right], \quad l \in \mathbb{Z} \qquad (3)$$

$$\rho = i\sqrt{\frac{\varepsilon_z}{\varepsilon_x \cos^2\varphi + \varepsilon_y \sin^2\varphi}}.$$

For simplicity, we neglect the reflection phases ($\Phi_\mathrm{a} = \Phi_\mathrm{m} = 0$). For each number $n$a and $n$m and fixed relation $w/d$, the wavevector $k_{a,m}$ is a function of a single parameter, $w$, which we represent by the solid and dashed curves in FigureS3b (labeled as "$n$a" and "$n$m", respectively).

Analogously to the case analyzed in the main text for the hyperbolic α-MoO$_3$ spectral range, for the elliptic spectral range we identified the specific FPRs corresponding to the peak positions in the $\delta R$ spectrum in FigureS3a. To that end we monitor the intersection between a horizontal line representing the inverse ribbon width (1.48μm) and the dispersion curves (Figure S3b). In the experimental spectrum, the resonant peaks at the frequencies matching $n$a = 2$n$m show the highest intensity. This can be explained by the 2 times larger PhP wavevectors of the M1 mode in the areas above the air gaps compared to those above the metal ribbons (due to the effective doubling of the flake thickness caused by the mirror effect of the metal). In contrast, in the hyperbolic range the modes are intrinsically different: while the regions above the air gaps support the M0 mode, the regions above the metal ribbons support the M1 mode. Consequently, the doubling effect is not observed in the hyperbolic frequency range. Note that the M0 mode in the flake above the air gaps in the elliptic range is not supported since the dielectric permittivities $\varepsilon_x$ and $\varepsilon_y$ are both positive.

FigureS3c shows the real part of the out-of-plane component of the vertical electric field across the ribbons as a function of frequency. The number of field oscillations across the ribbon coincides with the resonance numbers, $n$a and $n$m, labelling the resonance peak in Fig S3a. In FigureS3d, we illustrate the measured $\delta R$ spectra for $\varphi = 0°, 15°, 30°$, and $45°$, in which the positions of the resonances 2m and 4a agree well with the simulated spectra for all angles $\varphi$ (color plot in FigureS3e). Thus, we also see a strong dependence of the resonant frequencies on the rotation angles within the elliptic range, both in the experiment and theory.

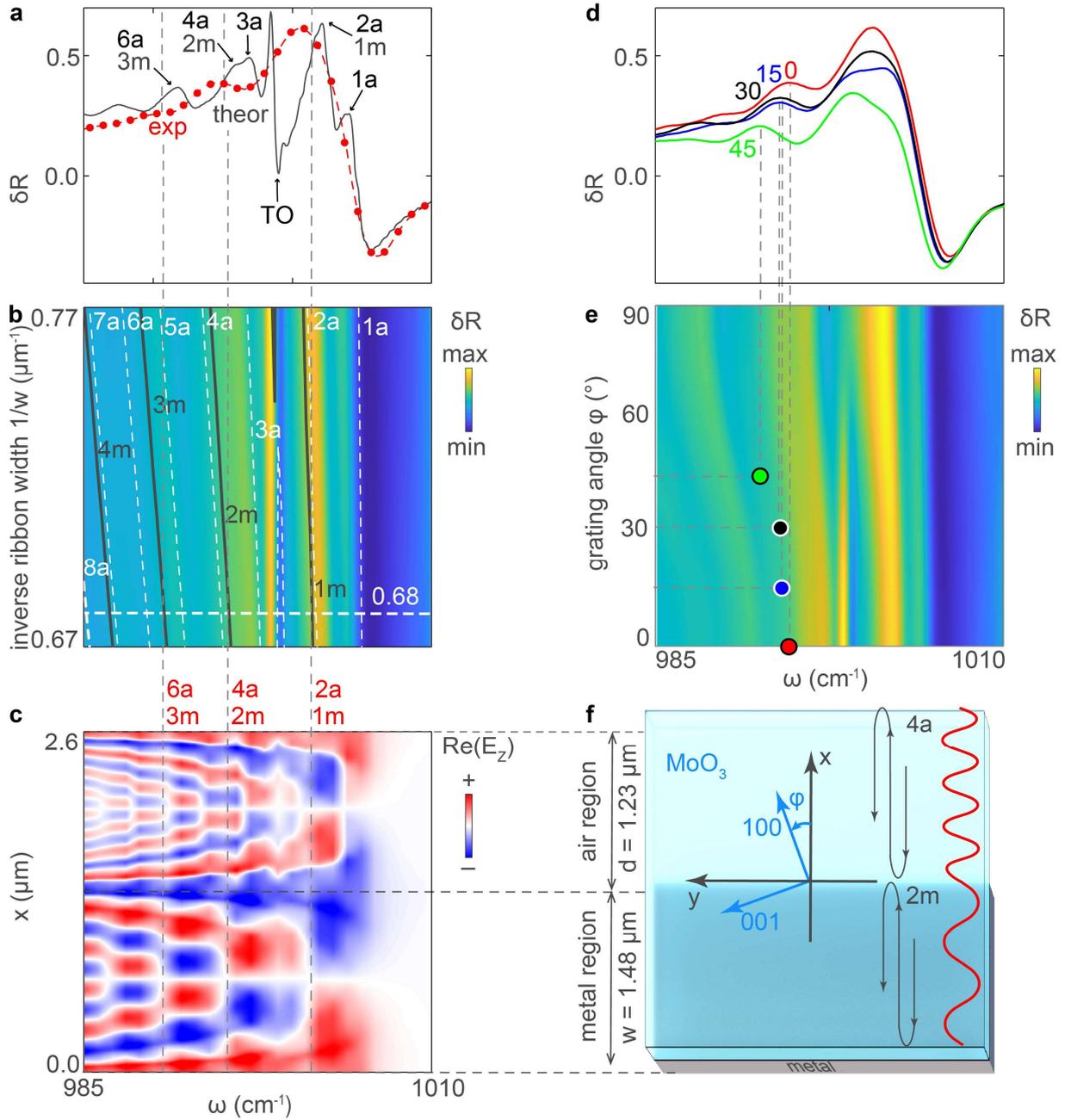

**Figure S3.** Analysis of the PhP FBRs in the elliptical range and their twist-tuning. a) Measured and simulated relative reflection spectra, $\delta R$ (dash red and solid grey curves, respectively), for a ribbon width $w = 1480$ nm, separation distance $d = 1230$ nm, and twist angle $\varphi = 0°$. b) Simulated $\delta R$ as a function of frequency $\omega$ and the inverse ribbon width, $1/w$, for a fixed ratio $w/d$. c) Simulated field distributions of the M1$_a$ and M1$_m$ PhP modes in the α-MoO$_3$/air and α-MoO$_3$/Au regions, respectively, as a function of $\omega$ and the $x$ coordinate (across the ribbons). d) Measured $\delta R$ spectra for the twist angles $\varphi = 0, 15, 30,$ and $45°$ (red, blue, black, and green, respectively). e Simulated relative reflection as a function of $\omega$ and $\varphi$. f) Schematics of the top view of one lattice unit cell.

# III. Dispersion Surface of the M1$_m$ PhP Mode above the Metal Ribbon Regions in the α-MoO$_3$ Hyperbolic Frequency Range

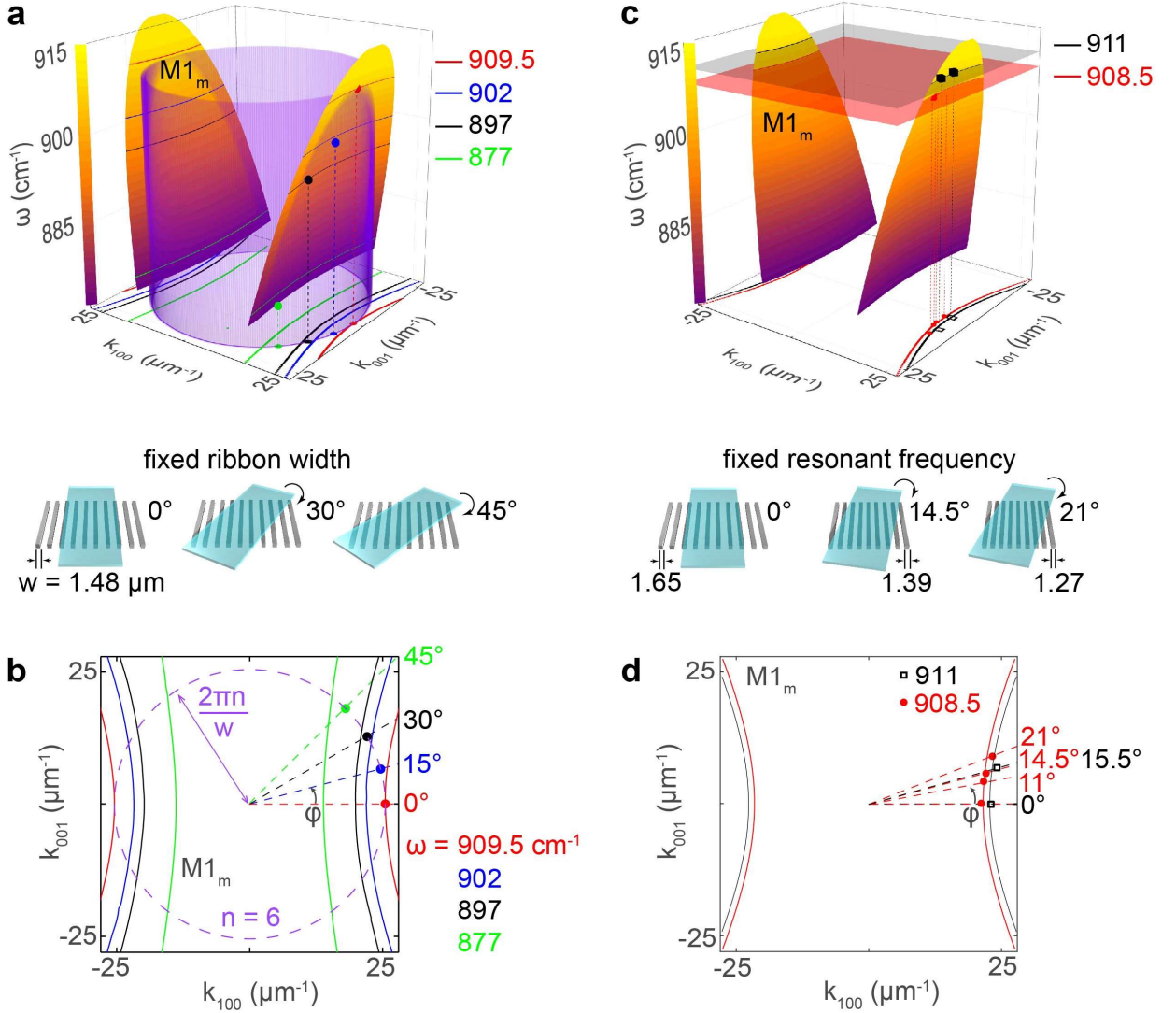

**Figure S4.** Probing the dispersion surface of the M1$_m$ PhP mode in the hyperbolic regime. a) Analytical dispersion surface of the M1$_m$ PhP mode crossed by a cylinder representing a constant momentum. The color dots represent the positions of the measured resonant peaks 6m in cylindrical coordinates (resonant frequency from Figure 2d, the inverse ribbon width of the grating, $2\pi n/w$, $n = 6$, and the twisting angle, $\varphi$) for the same Au grating twisted at different angles, $\varphi$. b,d) Isofrequency curves for the M1$_m$ mode at different frequencies, $\omega$. c) Analogous dispersion surface as in (a), but crossed by the planes of two fixed frequencies ($\omega = 908.5$, and 911 cm$^{-1}$). The red and black points mark the 6m peaks positions for Al gratings (Figure S2a and Figure S2b) in cylindrical coordinates (resonant frequencies, the inverse ribbon width of the grating, $2\pi n/w$, $n = 6$, and the twisting angle, $\varphi$). The thicknesses of the flakes in (a,b) and (c,d) are 110 and 127.5 nm, respectively.

In the main text probing the dispersion surface of the $M0_a$ PhPs mode (Figure 4) in the hyperbolic regime is performed for the 1a FBR in the area of the flake above air gaps. At the same time, the resonance 6m (the $M1_m$ mode) in the area of the flake above the metal ribbons can also be achieved at the same frequency as the resonance 1a for all twisting angles, $\varphi$. Thus, the same resonant peaks in the reflection spectra (Figure 2d) can be interpreted both as the 1a FBR of the $M0_a$ mode and the 6m FBR of the $M1_m$ mode. Here, we perform probing the dispersion surface of the $M1_m$ PhPs mode (Figure S4) in the hyperbolic regime on the example of the 6m FBR in the area of the flake above the metal ribbons. In Figure S4a the experimentally-measured positions of the FBRs are represented as green, black, blue and red points in cylindrical coordinates (the resonant frequency as a z coordinate, the inverse ribbon width, $2\pi n/w$, $n = 6$, as a radial coordinate and the twist angle, $\varphi$, as the polar angle). The purple cylinder with radius $2\pi n/w$, $n = 6$, represent the constant-momentum surface meaning that we perform the measurements for the same grating with fixed ribbon width, $w$, for all twist angles and frequencies, whereas the analytical dispersion surface of the $M1_m$ mode is represented by a hyperboloid. The projections of the points from Figure S4a on the ($k_x$, $k_y$) plane are positioned very closely to the intersection points between the analytical isofrequency curves (IFCs) for the corresponding frequencies and the cylinder projection on the ($k_x$, $k_y$) plane (Figure S4b).

An alternative probing of the PhP dispersion surface can be realized by constructing the cross-sections of the hyperboloid by constant-frequency planes (Figure S4c), i.e. IFCs corresponding to the resonant frequencies in Figure S2a,b. The projections of the experimentally-measured positions of the resonances, represented in cylindrical coordinates by red and black points in Figure S4c, fit well the IFCs.

Our results indicate that the dispersion surface of the $M1_m$ PhP mode in the hyperbolic frequency range can be experimentally reconstructed by measuring the resonance 6m in the region of the flake above the metal ribbons.

## IV. Dispersion Surfaces of the M1$_a$ and M1$_m$ PhP Modes above both the Air Gap and the Metal Ribbon Regions in the α-MoO$_3$ Elliptic Frequency Range

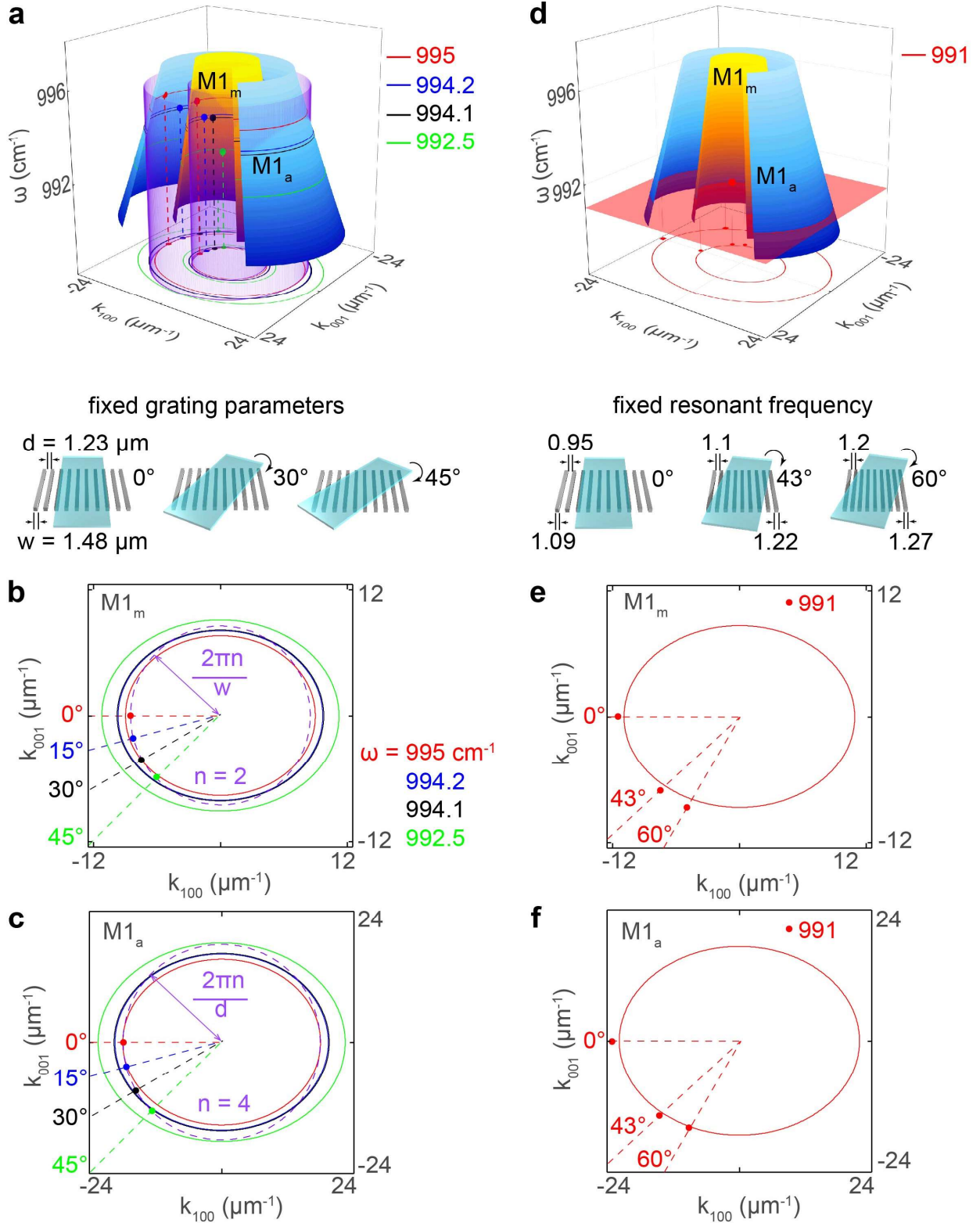

**Figure S5.** Probing the dispersion surfaces of the $M1_a$ and $M1_m$ PhP modes in the elliptic range. a) Analytical dispersion surfaces of the $M1_a$ and $M1_m$ PhP modes crossed by two cylinders representing constant momenta. The color dots mark the positions of the measured FPR peaks, 2m (near the yellow surface of the $M1_m$ PhP mode) and 4a (near the blue surface of the $M1_a$ PhP mode) in cylindrical coordinates (resonant frequency from Figure S3d; either the inverse ribbon width, $2\pi n/w$, $n = 2$, or the inverse air gap width, $2\pi n/d$, $n = 4$, for 2m and 4a resonances, respectively; and the twisting angle, $\varphi$) for the same Au grating twisted at different angles, $\varphi$. b,e) Isofrequency curves for the $M1_m$ mode at different frequencies, $\omega$. c,f) Isofrequency curves for the $M1_a$ mode at different frequencies, $\omega$. d) Analogous dispersion surface as in (a), but crossed by a plane of a fixed frequency, $\omega = 991 \text{cm}^{-1}$. The points mark the 2m and 4a FBR peak positions for Al gratings (resonant frequencies from Figure S2c, the inverse ribbon width of the grating, $2\pi n/w$, $n = 2$, and the inverse air gap width of the grating, $2\pi n/d$, $n = 4$, for 2m and 4a resonances, respectively, and the twisting angle, $\varphi$). The thicknesses of the flakes in (a,b,c) and (d,e,f) are 110 and 127.5 nm, respectively.

In analogy to the results shown in Figure 4 of the main text for probing the dispersion surface of $M0_a$ PhP mode in the hyperbolic range (performed for the 1a FBR emerging in the flake area above the air gaps), we represent in Figure S5 the results of the probing of the dispersion surfaces of the $M1_a$ and $M1_m$ PhP modes in the elliptic range for the 4a and 2m FPRs in the flake area above the air gaps and above the metal ribbons, respectively. The experimental data for Figure S5a-c was taken from Figure S3d, while the one for Figure S5d-f was taken from Figure S2c.

# V. Experimental Quality Factors of the Nanoresonators

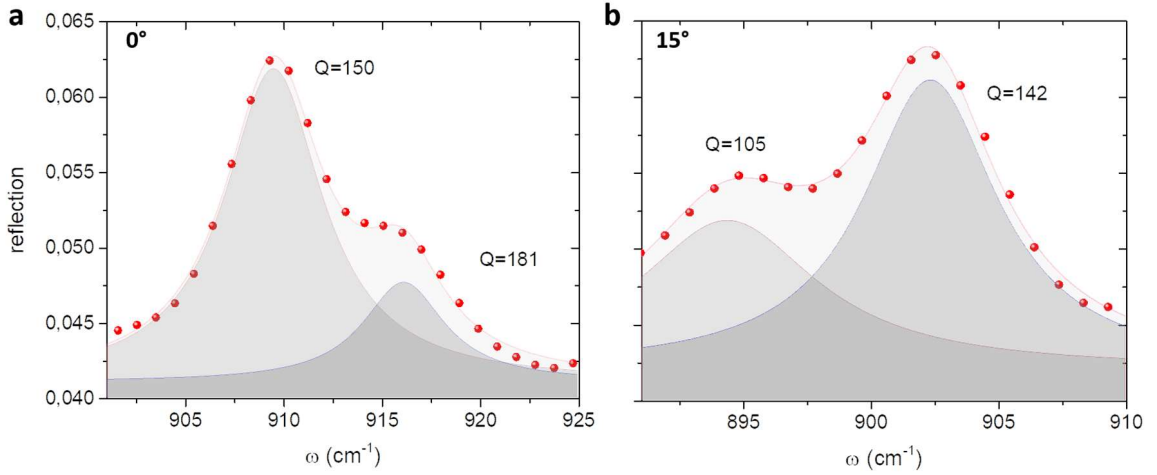

**Figure S6.** Experimental reflection spectra in the hyperbolic range for an α-MoO₃ flake placed on top of a metal grating with $w$ = 1.48 µm, $d$ = 1.23 µm. a) Rotation angle $\varphi = 0°$. b) Rotation angle $\varphi = 15°$.

To evaluate the quality factors ($Q$) of the nanoresonators, we choose resonant peaks corresponding to the 1a and 6m resonances in the α-MoO₃ hyperbolic range and the 4a and 2m resonances in the elliptic range. In order to extract $Q$, we fit the reflection spectra by two Lorentzian curves. The red symbols represent the experimental data measured, while the gray areas represent the fitted Lorentzian line shape to the experimental spectra (Figure S7). The quality factor can be defined as:

$$Q = \frac{\omega_{res}}{\Delta\omega} \tag{4}$$

where Q is the relation between the resonance frequency $\omega_{res}$ and the full width at half maximum $\Delta\omega$. We used the formula:

$$y = y_0 + \frac{2A}{\pi}\frac{\Delta\omega}{4(\omega-\omega_{res})^2+\Delta\omega^2}, \tag{5}$$

obtaining the following results for the hyperbolic range:

| Rotation angle φ (°) | $\omega_{res}$ (cm⁻¹) | | $\Delta\omega$ (cm⁻¹) | | Quality factor (Q) | |
|---|---|---|---|---|---|---|
| 0 | Peak 1 | 909.50 | Peak 1 | 6.05 | Peak 1 | 150 |
|   | Peak 2 | 916.14 | Peak 2 | 5.07 | Peak 2 | 181 |
| 15 | Peak 1 | 894.38 | Peak 1 | 8.56 | Peak 1 | 105 |
|    | Peak 2 | 902.35 | Peak 2 | 6.37 | Peak 2 | 142 |

| | | | |
|---|---|---|---|
| **30** | 896.97 | 8.86 | 101 |
| **45** | 876.25 | 6.56 | 134 |

Analogously, we fitted (in this case by one Lorentzian curve) the resonances measured in the elliptic range obtaining the following values:

| Rotation angle φ (°) | $\omega_{res}$ (cm$^{-1}$) | $\Delta\omega$ (cm$^{-1}$) | Quality factor (Q) |
|---|---|---|---|
| **0** | 995.16 | 7.07 | 141 |
| **15** | 994.76 | 12.66 | 79 |
| **30** | 994.56 | 14.56 | 68 |
| **45** | 992.76 | 4.99 | 199 |